\documentstyle[11pt]{aip}
\begin{document}

\title{DUMAND-II (Deep Underwater Muon and Neutrino Detector) \\
PROGRESS REPORT\\}

\author{ Kenneth K. Young \\
Dept. of Physics\\
University of Washington\\
Seattle, WA 98195\\
{\sl for The DUMAND Collaboration\cite{dumcollab} } }
\date{}

\maketitle

\begin{abstract}
The DUMAND II detector will search for
astronomical sources of high energy neutrinos.  Successful
deployment of the basic infrastructure, including the shore
cable, the underwater junction box, and an environmental module
was accomplished in December, 1993.  One optical module string
was also deployed and operated, logging data for about 10
hours.  The underwater cable was connected to the shore station
where we were able to successfully exercise system controls and
log further environmental data. After this time, water leaking
into the electronics control module for the deployed string
disabled the string electrical system.   The acquired data are
consistent with the expected rate of downgoing muons, and our
ability to reconstruct muons was demonstrated.  The measured
acoustical backgrounds are consistent with expectation, which
should allow acoustical detection of nearby PeV particle
cascades. The disabled string has been recovered and is
undergoing repairs ashore. We have identified the source of the
water leak and implemented additional testing and QC procedures
to ensure no repetition in our next deployment.  We will be ready
to deploy three strings and begin continuous data taking in late
1994 or early 1995.

\end{abstract}

\section*{INTRODUCTION}

The DUMAND Collaboration is building a neutrino observatory with
the aim of studying both diffuse and point sources of
astrophysical neutrinos in the TeV range.  Fluxes of charged
primary cosmic ray particles have been measured as a function of
energy and spectra have been presented to us at this conference
showing the characteristic knee and ankle features whose origins
are at present only subject to speculation.  Because neutrinos
are produced by some of the same processes that produce charged
particles, but are neutral and also only weakly  interacting,
they have the potential of revealing the spatial origins of the
cosmic rays.  Neutrino observations have helped elucidate the
origins of solar energy  production and the mechanism of
Supernova 1987A.  We expect DUMAND to elucidate the production of
high energy cosmic particles and their origins.

The Dumand II neutrino observatory\cite{roberts} is an array of
216 photomultiplier tubes deployed in nine vertical strings, in
an octagonal pattern with 40m sides and one string in the center.
 (DUMAND I designates a ship-suspended single prototype string
experiment successfully conducted in 1987.) The array will be
moored on the ocean floor at depth 4800m, 25 km from the Island
of Hawaii, and connected to a shore laboratory by a cable
combining electrical and fiber optic elements, terminating in an
underwater junction box. The
underwater site places no inherent limitation on possibilities
for future expansion of the detector.  DUMAND II when completed
will have an effective detection area of 20,000 m$^2$,
instrumenting a column of  water which has the height of the Eiffel
tower and its width at the base.

\section*{SUMMARY OF DUMAND II CONSTRUCTION PROGRESS}

The basic infrastructure of DUMAND, comprising the underwater
junction box, 30 km data and power cable to shore, and the shore
station facility are completed.   Environmental monitoring
equipment and the site-defining navigational sonar array have
been laid out and used in the 12/93 deployment operation.  One of
the optical module strings was deployed and used to record the
muon events.  Unfortunately, a hairline fracture in one of over
100 penetrators for the pressure vessels produced a small  water
leak.  Seawater eventually engulfed the electronics, disabling
further observation of muons after about 10 hours of operation.
The disabled string was remotely released and recovered at sea in
1/94, and returned to Honolulu for diagnosis and repair.  The
fault was found and ways to avoid future recurrences have been
identified.

Besides the refurbished first string, two further strings are
currently  undergoing final assembly and testing.  We  plan to
make extensive deep water tests of these three strings before
deploying them at the DUMAND site.   The earliest time that we
can obtain the ship and underwater vehicle resources needed to
carry out deployment and interconnection operations is around
December, 1994.

Signals from the optical modules are digitized locally providing
time of arrival (to 1 ns accuracy) and pulse height .  Signals from
the 24 OMs on each mooring are serialized and sent to
shore via an  optical fiber  at 0.5 GHz rate.  Technological
innovations in this system include the design and production of a
27 channel monolithic GaAs TDC chip with high
reliability and compact size.  This chip  has been implemented to
include all digitizer, buffer memory and multiplexing functions.
This system has  been built to cope with the background rate from
radioactivity in the water and bioluminescence and still
generate very little dead-time for recording cosmic events.  The
same optical fiber link will carry environmental and acoustical
ranging information which are used to measure the geometry of the
array.

The raw information is sent to the shore station 25 km away.  The
trigger system looks for patterns in  time, space and pulse
height of the OMs consistent with the passage of charged
particles through the array.  Events satisfying the trigger are
recorded for further off-line analysis.  Our studies of  the
trigger system predict that the system will be $>90$\% efficient for
events that penetrate the array from  the lower hemisphere.

Since  1992, DUMAND crews have been preparing the site and
testing underwater assembly  operations.  DUMAND II
requires a reasonably flat site with suitable soil properties.
The chosen site has
been marked with acoustical transponders which have been
accurately surveyed in geocentric coordinates.  The suitability
of this site was verified remotely by acoustical means, film
camera and video recordings; in addition, DUMAND personnel have
cruised the area in a manned submarine, the US Navy's {\it DSV
Sea Cliff}, to  be certain that the site is fairly flat and free
of any features that would interfere with a successful
deployment and operation.  We have verified the exceptional
clarity of the water.

After the deployment of the junction box with its single string
of OMs and the shore cable, each  successive string will be
moored in a ring at a radius of 40 m.  Strings will be connected
to the junction box by an umbilical cable and wet-mateable
electrical/fiber-optic connector.  Since this operation must be
carried out at a depth of 4800m, specialized underwater vehicles
must be used.  Using a mock junction box and string mooring, we
used the US Navy's Advanced Tethered Vehicle (ATV) to
maneuver within the array  and to carry out the connecting
operation.  The ATV successfully maneuvered into position,
unholstered the connector plug and cable, carried it in a
predetermined path and plugged it home in the socket.   We
proved that tethered vehicles (which are cheaper and more readily
available than manned submersibles) are capable of carrying out
this  operation with control from the surface.
The {\it DSV Sea Cliff} and ATV operations in late 1992 required
integration of our acoustical transponder system with surface GPS
navigation equipment and demonstrated that we can locate and work
at an ocean bottom site in a routine fashion.

We need to be able to point reconstructed muon tracks onto the
celestial sphere with an accuracy better than 1$^o$ (the
median angle between primary $\nu$ and secondary $\mu$ at 1
TeV). The global positioning satellite (GPS) system provides
accurate geographical coordinates for stations on the surface.
The transfer from the coordinates of satellite receiver antennas
on the surface to the underwater array is accomplished via an
acoustical positioning system of our own design and
construction\cite{berns93}.  In order to achieve reliable
positioning, we created a system that could  have a high signal
to noise ratio in spite of the long distances involved, the ocean
noise, and the multi-path interference of the sound waves.  The
system measures acoustical transit times with 10 $\mu$sec
precision and utilizes frequency modulated chirps and matched
filtering via DSPs to recover the signal.  We have  achieved 1 cm
accuracy in positioning in real time in short base line tests and
have positioned the site transponders to the accuracy of the GPS
system in the site survey.  In the final survey of the system, we
plan to use phase sensitive  techniques to survey the actual OMs
to an accuracy of $< 10$ cm.  The position  of the OMs will be
thereafter be monitored continually.

\section*{DEPLOYMENT OPERATIONS}

In December of 1994, a DUMAND scientific team and the crew of the
University of Washington oceanographic ship {\it RV Thomas G.
Thompson} were able to  successfully deploy all the elements of
one string and the infrastructure for eight more strings,
including the junction box, the environmental module, and the
shore cable.  Other  DUMAND scientific crews prepared the shore
station for operation.

The procedures for the lowering and cable laying operations had
been worked out in practice runs.  The cable laying equipment
was leased and mounted on the ship.  Last minute adjustment and
assembly of string components were completed, and final testing
was accomplished in refrigerated truck containers in a completely
connected configuration. The containers were then loaded onto the
{\it RV Thomas G. Thompson}, which was well
equipped to  handle all of this work.

At the time that DUMAND deployment operations began, the weather
was quite favorable and  the seas reasonably calm.  The practice
and planning paid off as we found that the complicated  operation
required to lower the string and junction box went  very well.
The navigation was excellent and the string and junction box were
landed well within the target areas.  Fig.~\ref{deploy} shows the
procedure used in  carrying out the deployment.  The string top
was attached to a sacrificial line and anchor which was  lowered
first.  The string followed, then came the junction box.   The
junction box was lowered on the shore cable to the bottom,
leaving the string in an arched configuration between the
junction box and the sacrificial anchor.  After touchdown the
cable was paid out and laid on the bottom as we headed for shore.
To avoid laying the  cable on the rocky shore which is pounded by
surf, divers threaded the cable through a previously prepared
slant-drilled tunnel which was bored from near the shore station
to an appropriate location offshore. By the end of the day, we had
hooked the shore cable up to power
and control cables and were able to  exercise controls on the
environmental module and acquire data.  This was an exciting day,
the culmination of years of planning and preparation.

\section*{RESULTS FROM THE DUMAND ARRAY}

We logged data from the DUMAND array as it was being lowered,
when it touched down on the bottom,  on shipboard during the
cable laying operation, and then from the shore station.  In all,
we recorded about 10 hours of  data.  The results are described
in the following sections.

We set a minimum threshold trigger of single photoelectron hits
on single OMs, in effect opening up the DAQ system to record
singles on all tubes.  We recorded data with this trigger for
about 10 hours.  Because of the 60 Khz singles rates, the DAQ
created quite a lot of dead time, primarily due to the time
required to dump a buffer of data to disk.  The live time
recorded was therefore approximately 2 minutes. These data were
then filtered offline for track candidates.  We have 10
candidate events with 6 or more OMs firing within a 100 nsec interval.
With a 60 Khz random singles rate, the expected number of events
from pure  chance is about $10^{-5}$.   The calculated rate of
downgoing muons is $2\times 10^6$/yr or 12 in the two minute
interval.  These data are thus well within expectation.
Fig.~\ref{timing}
shows the timing diagram for pulses from 7 OMs in our most
striking candidate.  The leading edges of the pulses are the
arrival times and the pulse width gives the time over threshold
(TOT)  which is proportional to the log of the integrated charge
collected by the OM.
The space-time hit
pattern roughly agrees with the hypothesis of a particle normal
to the string.  The brightness peaks at the  intersection point
and falls off rapidly in agreement with this hypothesis.  The
best fit hypothesis
is that there are two downgoing parallel particles. (Fig.~\ref{tracks}.)

Earlier investigations\cite{learned93} have suggested the possibility that a
very large volume and inexpensive detector of high energy neutrinos is
possible by acoustical detection.  The deposition of energy into
the water by the particles generates a low level characteristic
bipolar sound pulse with a frequency  range of  about 30 to 60
KHz.  Our simulation studies suggest that by using noise
cancellation and signal coherence techniques (ie, treating our
set of hydrophones as a phased array), we will be able to
systematically enhance noise rejection and detect high energy
particles.  The DUMAND II array is  equipped to observe
coincidences of OM and acoustical signals and this will provide
the first direct practical test of acoustical detection.

\section*{FUTURE PLANS}
Although the success of the DUMAND deployment was marred by the
failure of a single penetrator, we learnt  enough from the
limited period of live operation to be confident that we can
complete and operate the whole DUMAND array.  We also gained
confidence in the ability of our group to recover faulty
equipment from the sea, an essential task for long term
operation.   We are hoping that resources for the deployment of
the three strings can be available this winter.  The
capabilities of the three string array are discussed below.  A
demonstration of the viability of the three  string configuration
will allow us to complete the deployment of the following six
strings in the next year.

The capabilities of the full DUMAND II array have been reviewed
in previous reports\cite{icrc93}.  Here I will summarize  some of
the expected observations from the 3-string array.  Monte Carlo
simulations of the response of the  3-string array (Triad) will
have an effective detection area for muons above 3 TeV that
exceed previous and existing underground detectors.  The median
pointing accuracy at this energy will be about 3  degrees.  Thus
the Triad will be able to search for astronomical sources of very
high energy neutrinos at a greater level of sensitivity than has
so far been achieved in other experiments.  Our planned trigger
scheme, while keeping the trigger rate at a reasonable level, has
the unintended consequence of  a strong energy  dependence.  For
10 TeV muons, the Triad effective area will be 3100 m$^2$, which
scales roughly with  log(energy).  Using the neutrino fluxes
calculated by several authors for Active Galactic Nuclei (AGNs),
the following table of projected event rates is obtained:

\begin{table}
\caption{Expected AGN event rates in DUMAND for several models.}
\begin{tabular}[ph]{|l|l|}
\hline
Model\cite{HENA}	         & Event Rate (year$^{-1}$)\\
\hline
Biermann	       & 72 \\
Stecker et al	       & 97 \\
Sikora and Begeleman	& 71 \\
Protheroe and Szabo	& 21 \\
\hline
\end{tabular}
\end{table}

The rates calculated are the integral flux from all AGNs.  Except
for the  prediction of  Protheroe et al, all models predict
diffuse event rates which exceed the atmospheric background
rates.  Thus  the triad will have substantial capability for
detecting UHE cascades at distances of several hundred  meters,
and could provide the first evidence for diffuse AGN fluxes.

\section*{CONCLUSIONS}

We have demonstrated the viability of the DUMAND detector,
includ2ing successful deployment and operation of components
required for a large-scale underwater neutrino observatory.  The
December 1993 deployment operation was in effect a full up test
of the DUMAND design for hardware, software, system integration
and analysis procedures, and results were remarkably favorable,
given that DUMAND is one of the most complex oceanographic
projects ever undertaken.  Furthermore, tests with the acoustical
system of DUMAND show that we have the  capability of detecting
PeV cascades in the ocean with our present system of hydrophones.
We look forward to completing the full DUMAND II array.

We welcome you to find out more about DUMAND with text and
pictures accessed via the DUMAND Home Page on World Wide Web.
The URL address is
\noindent
\begin{verbatim}
http:\\web.phys.washington.edu/local_web/dumand/aaa_dumand_home.html
\end{verbatim}
You'll find colour photos, videos from the underwater camera, and
other DUMAND news.

Agencies providing
the funds for construction include the US DOE, HEP Division; the
Japanese Mombusho, from several funds;  the Swiss NSF; the US
NSF; all participating institutions, and the State of Hawaii.
We would particularly like to thank Vincent Z. Peterson and Syo
Tanaka, who retired recently, for their many contributions over
the years.

\begin{figure}[bh]
\vspace{2.0in}
\caption{Procedure used to deploy DUMAND string, junction box and shore
cable.}
\label{deploy}
\end{figure}

\begin{figure}
\vspace{3in}
\caption{Timing and TOT diagram for DUMAND event.}
\label{timing}
\end{figure}

\begin{figure}
\vspace{3in}
\caption{Best-fit hypothesis for DUMAND event: two downgoing muons. Event
was recorded before string was released to normal vertical
orientation; figure shows arched string configuration.}
\label{tracks}
\end{figure}

\end{document}